\documentclass[aps,prd,twocolumn,groupedaddress]{revtex4-2}
\usepackage{graphicx}
\usepackage{amsmath,amsfonts,amssymb}
\begin{document}

\title{Loop Special Relativity:\\ Kaluza-Klein area
metric as a line element for stringy events}
%\title{Kaluza-Klein mediated Loop Special Relativity:\\
%Area metric based line element for stringy events}

\author{Aharon Davidson}
\email{davidson@bgu.ac.il}
\homepage{www.bgu.ac.il/~davidson}
%\thanks{}
\affiliation{Physics Department, Ben-Gurion University
of the Negev, Beer-Sheva 84105, Israel}

\author{Nadav Barkai}
\email{barkain@gmail.com}
%\homepage{}
%\thanks{}
\affiliation{Physics Department, Ben-Gurion University
of the Negev, Beer-Sheva 84105, Israel}

\date{\today}

\begin{abstract}
Let a physical event constitute a simple loop
in spacetime.
This in turn calls for a generalized loop line element
(= distance$^2$ between two neighboring loops)
capable of restoring, at the shrinking loop limit,
the special relativistic line element (= distance$^2$
between the two neighboring center-of-masses,
respectively).
Sticking at first stage to a flat Euclidean/Minkowski
background, one is led to such a preliminary loop
line element, where the role of coordinates is
played by the oriented cross-sections projected
by the loop event.
Such cross-sections are generically center-of-mass
independent, unless (owing to a topological term)
the loop events are intrinsically wrapped around a
Kaluza-Klein like compact fifth dimension.
Serendipitously, it is the Kaluza-Klein ingredient
which, on top of its traditional assignments, is shown
to govern the extension of Pythagoras theorem to
loop space.
Associated with $M_4 \otimes S_1$ is then a
10-dim loop spacetime metric, whose 4-dim
center-of-mass core term is supplemented by a
6-dim Maxwell-style fine structure.
The imperative inclusion of a positive (say
Nambu-Goto) string tension within the framework
of Loop Special Relativity is fingerprinted by a low
periodicity breathing mode.
Nash global isometric embedding is conjectured to
play a major role in the construction of Loop General
Relativity.
\end{abstract}


\maketitle

\section{Rationale, setting and plan}

A mathematical event is by definition a point
in spacetime.
It marks, for example, the location $x^\mu$
of a classical point-like particle at some given instant. 
The square of the distance between two such
infinitesimally separated point-like events $x^\mu$
and $x^\mu+d x^\mu$, namely 
\begin{equation}
	ds^{2}=g_{\mu\nu}dx^{\mu}dx^{\nu} ~,
	\label{line}
\end{equation}
constitutes the special/general relativistic line
element, where $g_{\mu\nu}(x)$ is defined as the metric
tensor of the underlying flat/curved spacetime manifold.
The line element eq.(\ref{line}) has long been recognized
as the most fundamental geometrical tool in the service
of theoretical physics.

There is no compelling reason, however, why must
a classical physical event be inherently point-like, stripped
from any non-trivial (say stringy) micro structure. 
With this idea in mind, let $x^{\mu}(\sigma)$ define the
so-called simple, that is devoid of self intersections or
crossings, loop event (a refined definition of simplicity
will be given later).
As the $\sigma$-parameter varies from $0$ to $2\pi$,
it traces the path of a classical closed string at some
given instant.
The closed structure of the loop event, formulated by
\begin{equation}
	x^{\mu}(\sigma+2\pi)=x^{\mu}(\sigma)~,
	\label{2pi}
\end{equation}
gets manifested by means of the Fourier series expansion
\begin{equation}
	x^{\mu}(\sigma)=x^{\mu}_{cm}+\ell \xi^\mu (\sigma)
	=x^{\mu}_{cm}+
	\ell \sum_{n\neq 0}\xi^{\mu}_{n}~e^{in\sigma} ~.
	\label{Fourier}
\end{equation}
with $\xi_{-n}^\mu=\xi_{n}^{\mu\star}$.
The coefficient $\ell$ sets the loop length scale, leaving
the various $\xi^\mu_n$ dimensionless.

The immediate question now is the following:
Can one consistently construct, using a covariant
geometric formalism, a tenable loop line element
$\delta S$ to measure the generalized distance
between two such neighboring  loop configurations
$x^{\mu}(\sigma)$ and
$x^{\mu}(\sigma)+\delta x^{\mu}(\sigma)$?
The theoretical obstacle is threefold:
conceptual, technical, and furthermore dynamical.

% Fig1  %%%
\bigskip
\begin{figure}[h]
	\includegraphics[scale=0.25]{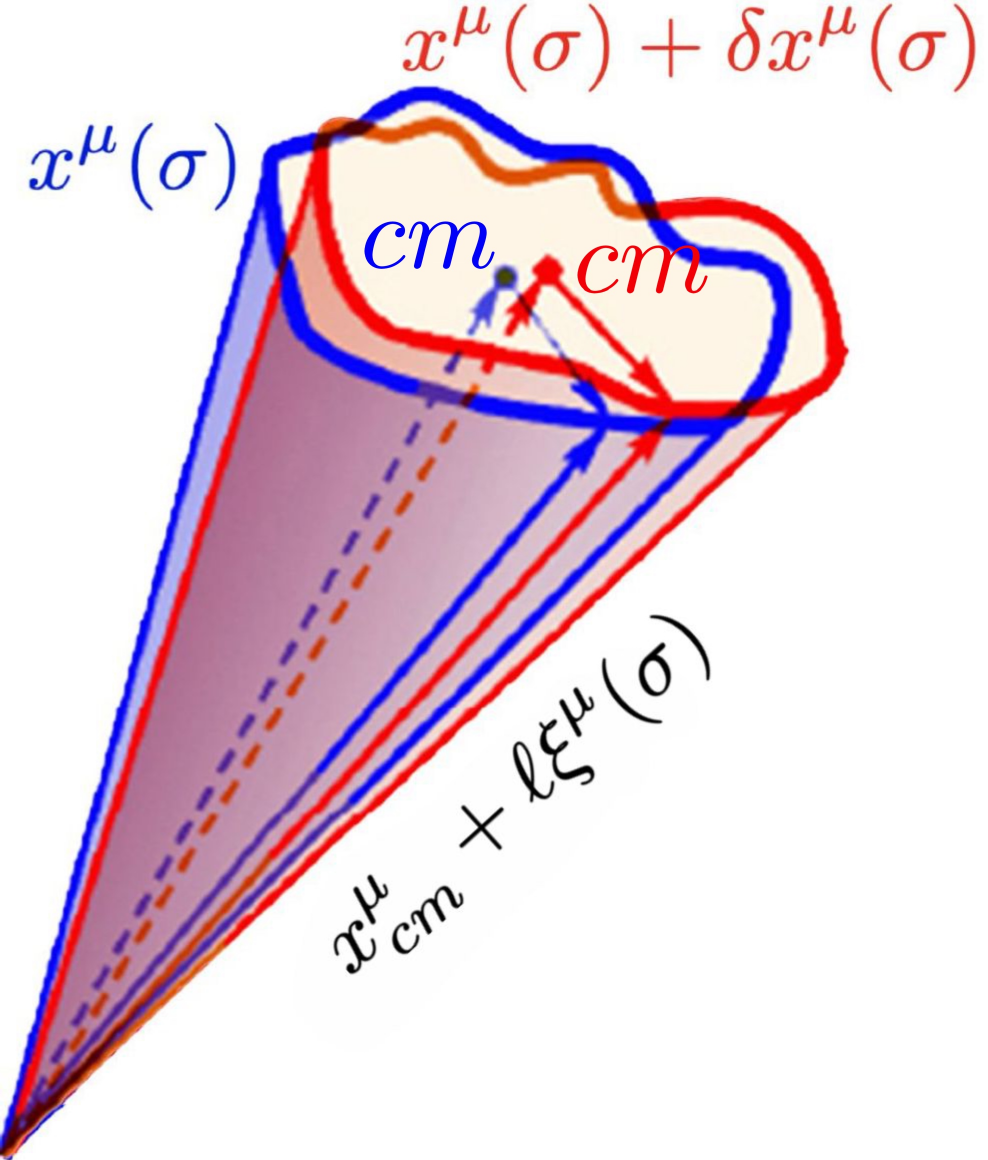}
	\caption{Two neighbouring loops in
	spacetime are mapped into two infinitely
	closed points in loop spacetime.
	A tenable loop line element $\delta S$ must
	then reproduce, at the shrinking loop limit
	$\ell\rightarrow 0$, the special/general relativistic
	line element measuring the distance between the
	two infinitely separated center-of-masses, respectively.}
	\label{fig1} 
\end{figure}

\noindent $\bullet$ On the conceptual level: 
While obviously dealing with non-local
(stringy) configurations, the loop line element must
nonetheless be local in loop spacetime.
To make geometrical sense out of such a requirement,
let each individual loop configuration be mapped into a
certain point in loop spacetime.
And the more so, two neighboring (infinitesimally deformed)
loops residing in spacetime must be mapped into two
infinitesimally separated points in loop spacetime.
This evidently calls for a tenable loop spacetime metric.
The precise identification of the entities which serve as
loop spacetime coordinates, the exact nature of the
loop-to-point mapping, and eventually the loop spacetime
metric, are to be presented and discussed (see Chapter 2).

\smallskip
\noindent  $\bullet$ On the technical level: 
There is a  crucial condition which, on self
consistency grounds, must be met at the shrinking
loop size limit.
As $\ell \rightarrow 0$, the loop line element
$\delta S$ must reproduce the special/general
relativistic line element eq.(\ref{line}), which will
then measure the distance between the two
associated infinitesimally closed center-of-masses,
respectively, that is
\begin{equation}
	\lim_{ \ell\rightarrow 0}\delta S^{2}\rightarrow
	g_{\mu\nu}dx_{cm}^{\mu}dx_{cm}^{\nu} ~.
	\label{SRlimit}
\end{equation}
Fulfilling this limit is a non-trivial technical task.
for example, the class of so-called area metrics
\cite{areametric}, which our preliminary loop line
element shares some common ingredients with,
fails to deliver in this respect.
The missing ingredient called to the rescue is
well known, albeit in a totally different area of
physics.
We refer to a compact fifth dimension a la
Kaluza-Klein \cite{KK}, which plays a novel
role in our discussion (see Chapter 3).

\smallskip
\noindent $\bullet$ On the dynamical level:
The rationale for replacing point-like events by 
loop events would not make any sense if the length
scale $\ell$ can grow arbitrarily large.
A dynamical physical mechanism, loop shape sensitive,
which would account for the natural shrinkage of loop
events, seems to be in order.
Such a mathematical service can be provided by
incorporating positive loop event tension a la
Nambu-Goto string theory \cite{NG}.
Once loop dynamics is introduced, one may
expects the length scale $\ell$ to eventually acquire
the Planck scale.
There is also room, and eventually a necessity,
as mentioned earlier, for a non-trivial spacetime
topology to enter the game.
In which case, relevant for our discussion (see
Chapter 3), loop events get wrapped around a
Kaluza-Klein-like cylinder.

A local realization of the loop line element idea
is expected to pave the way for a corresponding
Loop Special Relativity (LSR) theory. 
With this in mind, we first consider the case of a
flat spacetime which admits a Cartesian or pseudo
Cartesian metric $\eta_{ij}$, where $x^{i}$ itself,
rather than just the differential $dx^{i}$, transforms
as a vector.
Unfortunately, one immediately notices that the
desired special relativity limit eq.(\ref{SRlimit})
is generically not reachable.
Technically, it has to do with the geometrical fact that
the loop area is generically center of mass independent.
The remedy we offer requires a non-trivial topological
touch, and counter intuitively invokes the introduction
of an extra dimension.
To be specific, the loop event must be wrapped around
a spatial compact Kaluza-Klein like cylinder in order to
activate the explicit entrance of the otherwise
hidden center-of-mass coordinate into the loop metric. 
%Revised%
Put it differently, the only loops we are able to handle
via our approach are the ones wrapped around the fifth
dimension.
The physical role, if any, played by unwrapped loop
events is not discussed in this paper.
While our approach does not seem to have a direct
connection with the standard assignments \cite{KK}
of the Kaluza-Klein idea, it resembles some familiar
features.
For example, starting from an underlying
$M_4 \otimes S_1$ spacetime, the emerging loop
line element appears to be 10-dim, spanned by four
essential center-of mass coordinates, accompanied
by a 6-dim Maxwell style micro structure.

In Chapter 4 we pose the question whether 
string dynamics is imperative.
In other words,  is it necessary to accompany 
kinematical LSR by (say) dynamical Nambu-Goto,
and we give a few examples to support our
positive answer in a Euclidean background.
It turns out, however, see Chapter 5, that the
inclusion of Nambu-Goto action within the
framework of LSR has a unique fingerprint in
the Lorentzian background, namely a low
periodicity breathing mode.

The Loop Special Relativity (LSR) to Loop General
Relativity (LGR) generalization is still at large.
One idea in this direction would be to
invoke the Nash embedding formalism \cite{Nash},
later adopted by Regge-Teitelboim \cite{RT} in
its local isometric version within the framework
of geodesic brane gravity.
For example, given the constraint $x^2+y^2+z^2=1$,
curved $ds^2_2=d\theta^2+\sin^2 \theta d\phi^2$
can be trivially embedded within flat
$ds^2_3=dx^2+dy^2+dz^2$, so that the distance
between two loops residing on the $S_2$ sphere
gets translated (subject to the constraint) into the
distance between the two loops in the flat $E_3$
host.
A simple example is provided towards the end of
Chapter 3.
While every arbitrary curved space metric is Nash
embeddable, the embedding procedure itself has
several drawbacks:
The troubles are that
(i) The minimal embeddings are in general
local, not global,
(ii) The embedding is not necessarily unique, and
(iii) The number of embedding dimensions is very
much case dependent.
Another idea towards constructing LGR calls
for geometric gauge invariance of the second type
\cite{Barkai}.
Regretfully, this line of research lies beyond the
scope of the present paper.
The same holds or Rosen's bi-metric approach
\cite{Rosen}.

LSR does not seem to show, at least at this stage,
any compelling connection with Loop Quantum
Gravity (LQG) \cite{LQG}.
Still, encouraged by the 4-dim center of mass
resurrection in the loop area formalism, hereby
considered as the fingerprint of a compact 5-th
dimension (see Chapter 3), establishing such a
bridge is certainly welcome.

\section{Preliminary loop line element} 

Let our starting point be a loop drawn in
a flat plane characterized by the Euclidean metric
\begin{equation}
	ds^2=dx_1^2+dx_2^2 ~.
	\label{2flat}
\end{equation}
The area enclosed by a loop is given by
\begin{equation}
	A=\frac{1}{2}\oint (x_1 dx_2-x_2 dx_1)
	=\frac{1}{2}\int_0^{2\pi}r^2d\sigma ~.
	\label{area}
\end{equation}
Being a global quantity, the enclosed area $A$
is not sensitive to the fine local structure of the
 loop configuration. 
The mapping from the 2-dim $\{x_1,x_2\}$ plane
onto the 1-dim $A$-axis is thus not one-to-one.
And most importantly, to be regarded a momentary
drawback for our purposes, it has nothing to do
with the location of the center of mass of the loop.

Eq.(\ref{area}) can be easily generalized for the case
of a  loop residing within a larger flat space
(or spacetime) equipped
with a Cartesian (or pseudo Cartesian) coordinate
system.
In which case, the projected areas $A^{ij}$ are given by
\begin{equation}
	A^{ij}=\frac{1}{2}
	\oint (x^i dx^j-x^j dx^i)
	=\frac{1}{2}\int_0^{2\pi}
	(x^i x^{\prime j}
	-x^j x^{\prime i})d\sigma ~,
	\label{Aij}
\end{equation}
where $\displaystyle{f^{\prime}\equiv\frac{\partial f}
{\partial \sigma}}$.
For an $n$-dim spacetime, there are now
$\frac{1}{2}n(n-1)$ such projected areas,
one per each pair of spacetime indices.
It should be emphasized that it is only for the case
of a flat rectangular space (or spacetime) that
(i) $x^i$  transforms as a vector itself, to be
contrasted with the differential $dx^i$ which
always does, and
(ii) Owing to the global nature of the associated
Lorentz transformations, the integration over a tensor
is mathematically permissible.
In other words, $A^{ij}$ as given by eq.(\ref{Aij})
constitutes a rank-2 anti-symmetric tensor in
flat spacetime.

Now, for any given $\sigma$ (keeping $\sigma$ untouched),
consider a loop variation
\begin{equation}
	x^i (\sigma)\rightarrow
	x^i (\sigma)+\delta x^i (\sigma) ~.
\end{equation}
Following Euler-Lagrange, and subject to the periodicity
condition eq.(\ref{2pi}), we find
\begin{equation}
	\delta A^{ij}=
	\oint ( \delta x^i dx^j-\delta x^j dx^i) ~.
\end{equation}
The increment $\delta x^i (\sigma)$ can be controlled
by some parameter $\tau$.
In which case, we have $\delta x^i =\dot{x}^{i} d\tau$,
where $\displaystyle{\dot{f}\equiv\frac{\partial f}{\partial \tau}}$.
In turn, the projected areas $A^{ij}$ get shifted by
\begin{equation}
	\delta A^{ij}=d\tau \int_0^{2\pi} 
	(\dot {x}^i x^{\prime j}
	-\dot {x}^j x^{\prime i})d\sigma ~.
	\label{Aflat}
\end{equation}
Note that, owing to the built-in $i\leftrightarrow j$
anti-symmetry, the same result would have been
obtained had we started from the more general
expression
$\delta x^i =\dot{x}^{i} d\tau+x^{\prime i} d\sigma$.

While the first derivative $\displaystyle{\frac{dA^{ij}}{dt}}$,
involving the troublesome contour integration, behaves as
a tensor solely in flat spacetime, it is the second derivative
\begin{equation}
	\frac{d^2 A^{ij}}{d\tau d\sigma}=
	\dot {x}^i x^{\prime j}-\dot {x}^j x^{\prime i}
\end{equation}
which appears to constitute a legitimate tensor even in
curved spacetime.
This may be the point to start from when attempting
to eventually generalize LSR into LGR. %Revised%

The anti-symmetry of the oriented cross sections
is a fundamental feature, and does not depend on
the structure of the spacetime metric.
The more so,
\begin{equation}
	\epsilon^{\alpha\beta}x^i_{,\alpha} x^j_{,\beta}
	\quad\quad
	\{\alpha,\beta\}=\tau,\sigma
\end{equation}
serves as a set of world sheet scalar densities
associated with the world sheet induced metric
$\gamma_{\alpha\beta}=
\eta_{ij}x^i_{,\alpha} x^j_{,\beta}$.
The integrant within eq.(\ref{Aflat})
is thus re-parametrization invariant.
We note in passing that performing a proper
re-parametrization transformation 
\begin{equation}
	\Bigl\{
	\begin{array}{lcc}
	 \tau\rightarrow \tilde{\tau}(\tau,\sigma)
	 =T(\tau) &&  \vspace{3pt} \\
	\sigma\rightarrow \tilde{\sigma}(\tau,\sigma)
	=\sigma+\Sigma(\tau)  &&   
	\end{array} ~,
\end{equation}
prior to the integration, a transformation which
fully respects the
$\Delta \tilde{\sigma}=\Delta\sigma=2\pi$, 
 loop event periodicity, is 
equivalent to posteriori gauge fixing
$\tau\rightarrow T(\tau)$.

Once curvilinear coordinates are being used, while
the integration over a tensor is apparently forbidden,
there is a simple way out.
The trick is to invoke the Vierbein formalism,
where the curvilinear flat spacetime metric can be
written in the form
\begin{equation}
	g_{\mu\nu}=\eta_{ij}
	V^i_{,\mu} (x) V^j_{,\nu}(x) ~,
\end{equation}
notably involving total derivative Vierbeins.
Eq.(\ref{Aflat}) can then be easily generalized into
\begin{eqnarray}
	&\delta A^{ij}=\displaystyle{d\tau \int_0^{2\pi}
	\epsilon^{\alpha\beta} V^i_{,\mu} V^j_{,\nu}  
	x^\mu_{,\alpha} x^\nu_{,\beta} d\sigma}& \nonumber \\ 
	&=\displaystyle{d\tau \int_0^{2\pi}
	\epsilon^{\alpha\beta} V^i_{,\alpha} V^j_{,\beta} 
	d\sigma} ~& ~,
	\label{Acurvi}
\end{eqnarray}
and one is back to the original case, only with
the rectangular coordinates $V^i(x)$ replacing
the curvilinear coordinates $x^\mu$.

The situation is totally different, however,  for
a curved spacetime characterized by a non-vanishing
Riemann tensor).
Having in mind eq.(\ref{Acurvi}), one may
prematurely expect that associated with a
curved spacetime metric
\begin{equation}
	g_{\mu\nu}(x)=\eta_{ij}
	V^i_\mu (x) V^j_\nu (x)~,
\end{equation}
where the Vierbeins have now matured into full
gauge fields, are projected loop area increments
of the form
\begin{equation}
	\delta A^{ij}=d\tau \int_0^{2\pi}
	\epsilon^{\alpha\beta} V^i_\mu V^j_\nu  
	x^\mu_{,\alpha} x^\nu_{,\beta} d\sigma ~.
	\label{Awrong}
\end{equation}
Unfortunately, with the exception of the 2-dim
case, this formula does not make sense mathematically.
The problem is that it is impossible to erect any
single coordinate system that is locally inertial
everywhere, unless the spacetime continuum is
flat.
In other words, integration over a Lorentz tensor
does not make then any sense.
It is only in 2-dim, owing to the corresponding
rank-2 Levi-Civita symbol $\epsilon_{ij}$, that
$\delta A^{12}=\frac{1}{2}\epsilon_{ij}\delta A^{ij}$,
as defined by eq.(\ref{Awrong}), happens to be a
Lorentz scalar.
From this point on, on both pedagogical and
simplicity grounds, while still lacking a proper
formula for $\delta A^{\mu\nu}$ in a curved
background, and without invoking either
(i) gauge invariance of the second type, and/or
Regge-Teitelboim like embedding technique as
potential remedies, we return to the comfortable
choice of a Cartesian (or pseudo Cartesian)
spacetime.

\begin{figure}[h]
	\center
	\includegraphics[scale=0.25]{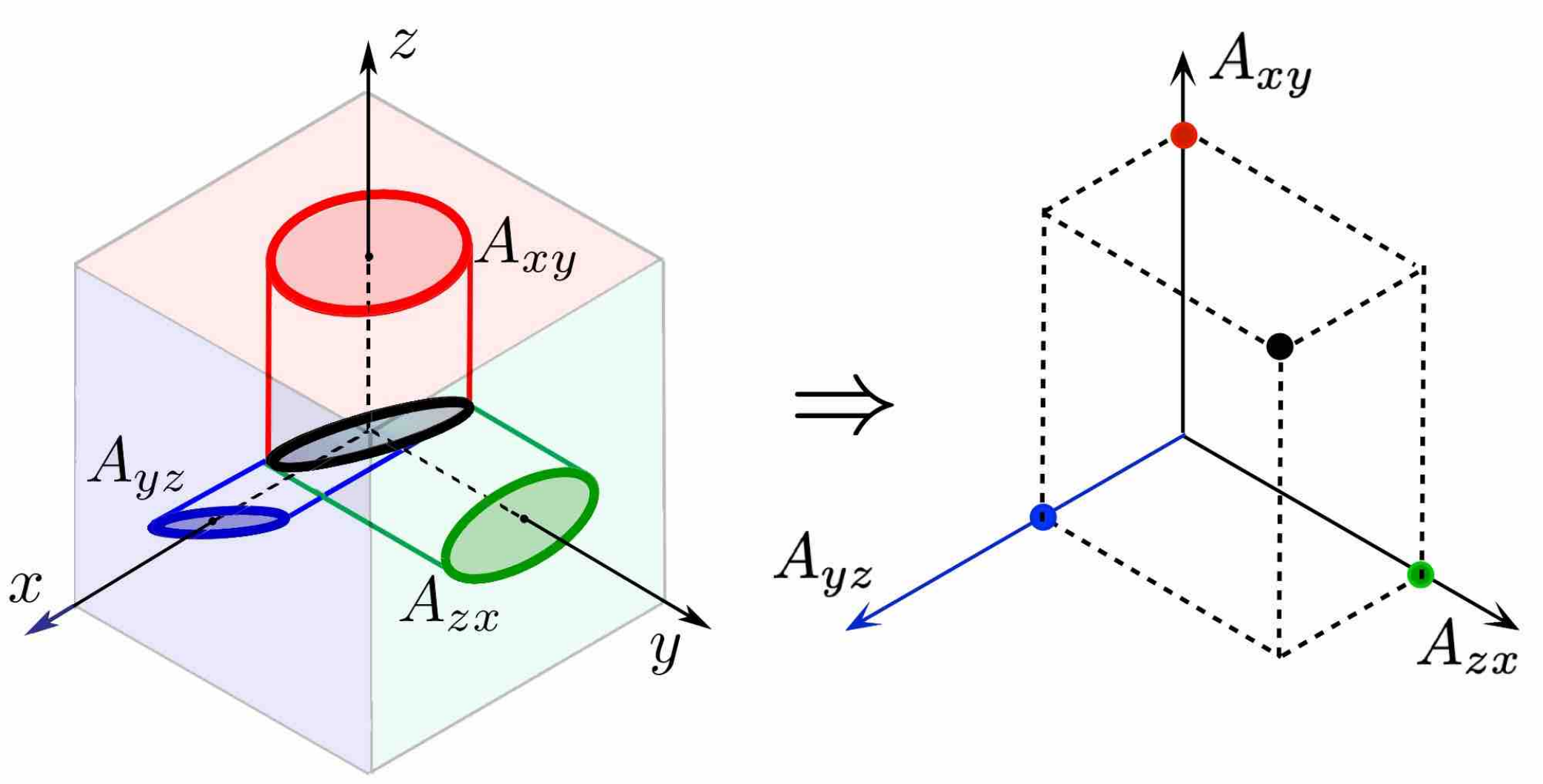}
	\caption{Associated with every simple 
	loop event (black) residing in an $n$-dim flat
	space or spacetime, there are $\frac{1}{2}n(n-1)$
	simple projected areas (red, blue, green,...), thereby
	mapping a loop event into a point
	$\{A_{xy},A_{yz}, A_{yz},...   \}$ in loop space.}
	\label{fig2} 
\end{figure}

Associated with each  loop residing
in an $n$-dim spacetime there is a
corresponding point in the so-called
loop spacetime, where the
$\frac{1}{2}n(n-1)$ independent projected
areas $A^{ij}$ play the role of coordinates
(see Fig.\ref{fig2}).
Two such  loops are then mapped into
two points in loop spacetime, and in principle,
provided the corresponding loop metric
is specified, a geodesic trajectory can be drawn.
If the two points are infinitesimally closed, we
pay tribute to the anti-symmetric structure of
$A^{ij}$, imitate the general relativistic structure
of the Maxwell kinetic term, and accordingly
define the scalar loop line element
\begin{equation}
	\boxed{
	\delta S^2=
	\frac{\eta_{ik}\eta_{jl}}{16\pi^2\ell^2}
	\delta A^{ij} \delta A^{kl}}~,
	\label{loopline}
\end{equation} 
properly normalized for future assignments.
Note that for the special case $n=4$,
there exists the option of supplementing
eq.(\ref{loopline}), or even replace it, by the
dual term proportional to
$\epsilon_{ijkl}\delta A^{ij} \delta A^{kl}$.
Eq.(\ref{loopline}) falls into the category of area
metrices \cite{areametric}.
Note in passing that areas also play an important
role in Regge \cite{Regge} calculus, and oriented
areas enclosed by string constitute a vital part of
Clifford space metric \cite{Clifford}.

% Fig2  %%%

Unfortunately, the special relativistic limit eq.(\ref{SRlimit})
is still unattainable at this stage.
Substituting the Fourier expansion eq.(\ref{Fourier})
into eq.(\ref{Aflat}), we obtain
\begin{equation}
	\delta A^{ij}= \ell^2
	\int_0^{2\pi} (\dot{\xi}^i \xi^{\prime j}
	-\dot{\xi}^j \xi^{\prime i}) d\tau d\sigma ~,
	\label{dAij}
\end{equation}
and immediately notice that the center of mass
$x^i_{cm}(\tau)$ and its $\tau$-derivative
$\dot x^i_{cm}(\tau)$ do not enter the game.
This poses a major drawback, as all point-like
($=$ shrinking loop) events become practically
indistinguishable, and pile at the origin.
Until the missing ingredient is found, and the center
of mass comes out of hiding, eq.(\ref{loopline}) has
to be regarded incomplete.

%%%%%%%%%%%%%%%%%%%%
\section{Kaluza-Klein to the rescue}
\subsection{Center of Mass resurrection}

Counter intuitively, the remedy comes from topology.
The idea is to supplement spacetime by an extra
Kaluza-Klein like closed dimension.
Nothing to do though with the original Kaluza-Klein
idea.
From some reason soon to be clarified, this extra
dimension must be space-like.
On historical grounds we generically refer to such
an extra dimension as $x^5$, and consider loop
events wrapped around this cylindrical fifth dimension.
The previously introduced spacetime coordinates
$x^{\mu}$ are now accompanied by a new $x^{5}$,
subject to the periodicity condition
\begin{equation}
	\Delta x^5=2\pi  R ~.
\end{equation}
The game changer element is then the presence
of the topological term $R\sigma$ in the  modified
Fourier expansion.
To be contrasted with eq.(\ref{Fourier}), we now have
\begin{equation}
	x^{5}(\sigma)=x^{5}_{cm}+R\sigma+
	\ell \xi^{5}(\sigma) ~.
\end{equation}
where as usual $\xi^5 (\sigma)=
\displaystyle{\sum_{n\neq 0}}\xi^{5}_{n}~e^{in\sigma}$.

Owing to the slight yet significant modification
in the corresponding partial derivative expansions,
that is
\begin{equation}
	\Bigl\{
	\begin{array}{lcc}
		\dot x^{5}=
		\dot x^{5}_{cm}+\ell \dot \xi^5 && 
		\vspace{4 pt} \\
		x^{\prime 5}=R+\ell \xi^{\prime 5} &&   
	\end{array}
\end{equation}
in comparison with the former
\begin{equation}
	\Bigl\{
	\begin{array}{lcc}
		\dot x^{i}=\dot x^{i}_{cm}+\ell 
	\dot \xi^{i} &&  \vspace{4 pt} \\
		x^{\prime i}=\ell  \xi^{\prime i}   &&   
	\end{array}~,
\end{equation}
the projected areas split into two distinguishable
categories.
While, as before, the center of mass derivative
$\dot x^{i}_{cm}$ is still absent from
\begin{equation}
	\frac{\delta A^{ij}}{2\pi }=
	 \left(\frac{\ell^2}{2\pi}
	\int_0^{2\pi} (\dot{\xi}^i \xi^{\prime j}
	-\dot{\xi}^j \xi^{\prime i}) d\sigma
	\right)d\tau ~,
	\label{dAij2}
\end{equation}
it makes its resurrection via
\begin{equation}
	\frac{\delta A^{i5}}{2\pi }=
	 \left(R \dot x^{i}_{cm}
	+\frac{\ell^2}{2\pi}
	\int_0^{2\pi} (\dot{\xi}^i \xi^{\prime 5}
	-\dot{\xi}^5 \xi^{\prime i}) d\sigma
	\right)d\tau ~.
	\label{Ai5}
\end{equation}
Notably, as expected, and in contrast with the
presence of $\dot x^{i}_{cm}$, irrelevant
$\dot x^{5}_{cm}$ stays completely out of the game.

Equipped with eqs.(\ref{dAij2},\ref{Ai5}), one can now
reconstruct the loop line element eq.(\ref{loopline}).
In its modified version, reflecting the underlying
$M_n \otimes S_1$ spacetime,
it takes the final form
\begin{equation}
	\boxed{
	\delta S^2=\frac{1}{8\pi^2R^2}
	\left(\eta_{ij}\eta_{55}
	\delta A^{i5} \delta A^{j5}+
	\frac{1}{2}\eta_{ik}\eta_{jl}
	\delta A^{ij} \delta A^{kl}\right)}
	\label{newloopline}
\end{equation}
Here, by requiring $\eta_{ij}\eta_{55}=\eta_{ij}$
on consistency grounds, in order  to account for
the mandatory $\eta_{ij}x_{cm}^i x_{cm}^j$ term,
we are finally led to the tenable signature choice 
\begin{equation}
	\eta_{55}=+1 ~,
\end{equation}
a posteriori justifying the space-like nature of $x^5$
(nothing to do with the original Kaluza-Klein theory).
On the other hand, given the underlying $\eta_{ij}$
metric, the coefficients $\eta_{ik}\eta_{jl}$ are
uniquely fixed.

\subsection{Pythagoras theorem in loop space}

\begin{figure}[h]
	\center
	\includegraphics[scale=0.23]{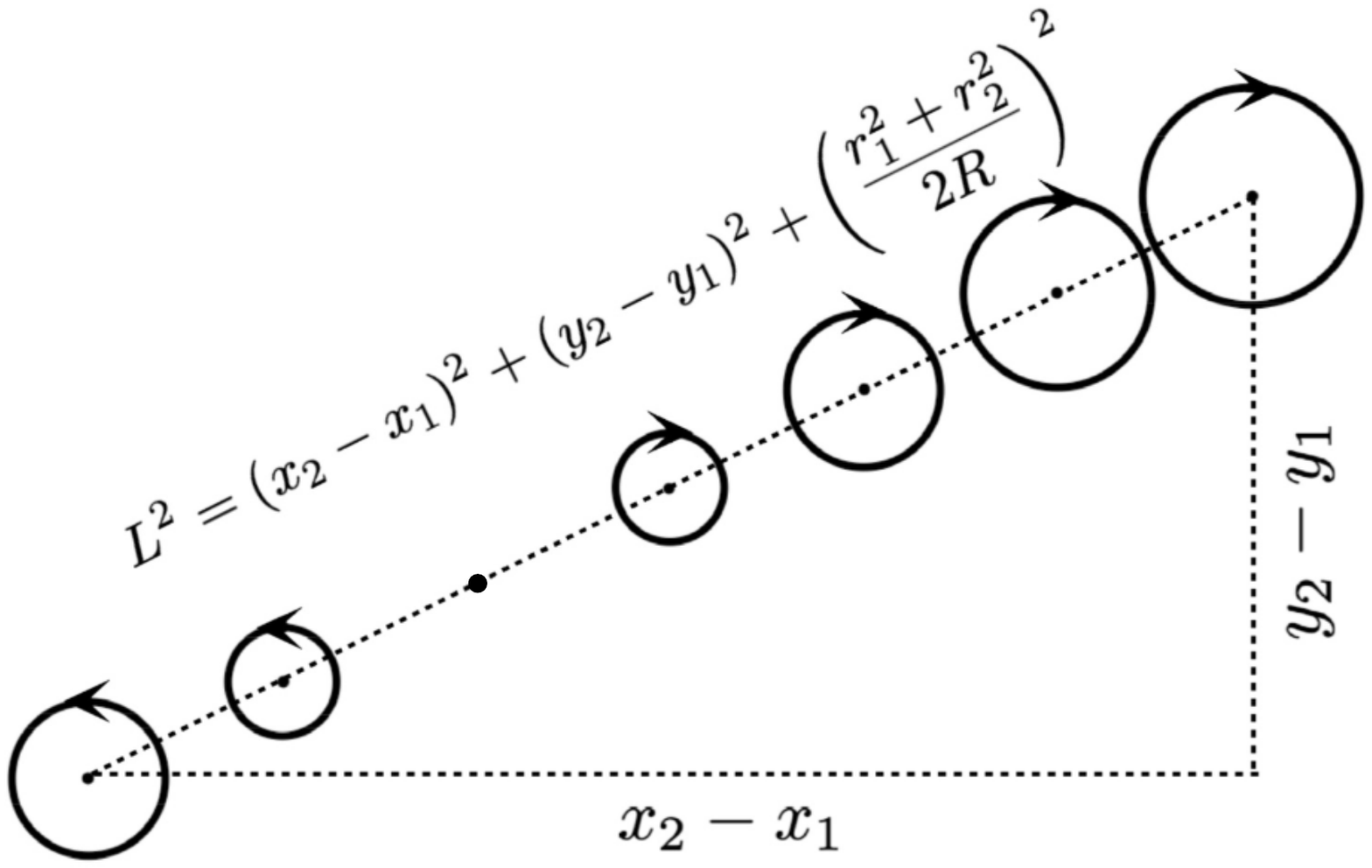}
	\caption{Generalized Pythagoras theorem:
	Let $L$  denote the geodesic distance between
	two loops, demonstrated here for circles of
	opposite sides, so that
	$A_2-A_1=\pi(r_2^2-(-r_1^2))$.
	While the shape of the loop stays unconstrained
	in the absence of loop dynamics, the center of
	mass location, as well as the loop area, evolve
	linearly.}
	\label{fig3} 
\end{figure}

The special case we now discuss in detail is the
simplest, yet the most fundamental case in hands.
And as such, has been moved from the Appendix
level to the main body of our paper.
To be specific, let us calculate the distance
between two arbitrary loops residing in a flat
2-dim plane 
\begin{equation}
	ds^2=dx^2+dy^2 ~.
\end{equation}
Following our prescription, our first step is to add
the Kaluza-Klein ingredient to the game, and deal
with a larger yet flat 3-dim space governed by
\begin{equation}
	d\bar s^2=ds^2+dx_5^2 ~,
	\quad \Delta x_5=2\pi R ~.
\end{equation}
On simplicity and pedagogical grounds, we consider
the evolution (parametrized by $\lambda$) of a circular
loop.
The loop is of radius $r(\lambda)$, is centered at
$\{x_{cm}(\lambda),y_{cm}(\lambda)\}$
\begin{equation}
	\begin{array}{lll}
		& x(\lambda,\sigma)=
		x_{cm}(\lambda)+r(\lambda)\cos\sigma &  
		\vspace{3pt} \\
		&y(\lambda,\sigma)=
		 y_{cm}(\lambda)+r(\lambda)\sin\sigma &
		 \vspace{3pt}  \\ 
		& x_5(\lambda,\sigma)=
		x_{5cm}(\lambda)+R\sigma & 
	\end{array} ~,
\end{equation}
and is furthermore wrapped, precisely once at
this stage, around the Kaluza-Klein cylinder.
The three projected areas are given explicitly by
\begin{eqnarray}
	&&\frac{dA^{xy}}{d\lambda}=
	2\pi r \frac{dr}{d\lambda} ~, \\
	&&\frac{dA^{x5}}{d\lambda}=
	2\pi R \frac{dx_{cm}}{d\lambda} ~, \\
	&&\frac{dA^{y5}}{d\lambda}=
	2\pi R \frac{dy_{cm}}{d\lambda} ~.
\end{eqnarray}
Substituting the latter into eq.(\ref{newloopline}),
we immediately find out that
\begin{equation}
	dS^2=dx_{cm}^2+dy_{cm}^2+
	\frac{r^2}{R^2}dr^2 ~,
\end{equation}
and recalling the loop area
$A^{xy}=\pi r^2$,
brings us to the final Pythagoras formula
\begin{equation}
	\boxed{dS^2=ds_{cm}^2+
	\left(\frac{dA}{2\pi R}\right)^2}
	\label{Pythagoras}
\end{equation}
We have thus reached a flat 3-dim loop space,
with $\displaystyle{\frac{A}{2\pi R}}$
serving as a third dimension.
It is an open dimension, to be contrasted
with the compact nature of $x^5$, thus mimicking
the role of $z_{cm}$.
Altogether (see Fig.\ref{fig3}),
the geodesic distance $L$ between
two loops, of areas $A_1$ and $A_2$, centered
at $\{x_1,y_1\}$ and $\{x_2,y_2\}$ respectively,
is given by
\begin{equation}
	L^2=(x_2-x_1)^2+(y_2-y_1)^2+
	\left(\frac{A_2-A_1}{2\pi R}\right)^2 ~.
\end{equation}

One may now wonder under what circumstances
is the (area difference)$^2$ term negligible?
The trivial answer would be when $A_1=A_2$,
the case of a fixed loop area (the rigid
loop case obviously included).
The generic case calls for $A^{xy}\sim \ell^2$ to
pick up the only length scale floating around.
In general, however, as far as the geometry is
concerned, the area difference $A_1-A_2$ stays
at this stage unbounded.

% Fig3  %%%

\subsection{Light cone in loop spacetime}

Had one started from a Lorentzian spacetime
$M_2$ enriched by a compact ($\Delta x_5=2\pi R$)
fifth dimension
\begin{equation}
	d\bar{s}^2=-dt^2+dz^2+dx_5^2 ~,
\end{equation}
and analogously considered the simple
$\tau$-evolution 
\begin{equation}
	\begin{array}{lll}
		& t(\tau,\sigma)=t_{cm}(\tau)
		+a(\tau)\cos\sigma&  
		\vspace{3pt} \\
		& z(\tau,\sigma)=
		z_{cm}(\tau)+b(\tau)\sin\sigma &
		\vspace{3pt} \\
		 & x_5(\tau,\sigma)=
		 x_{5cm}(\tau)+R\sigma & 
	\end{array} ~,
\end{equation}
he would have once again ended up with the
fundamental eq.(\ref{Pythagoras}), only with
$A^{tz}=\pi ab$.
Note that $A^{ij}$ is not sensitive to the
time/spacelike nature of the $ij$-indices.

Altogether, as long as loop dynamics is not
switched on, the kinematical evolution of a
loop event, residing in $M_2$ and being
wrapped around $S_1$, is described by means
of the geodesic evolution of a point-like event
in flat $M_3$.
While the Poincare-like symmetry associated with
the loop spacetime metric
\begin{equation}
	dS^2=-dt_{cm}^2+dz_{cm}^2-
	\left(\frac{dA}{2\pi R}\right)^2
\end{equation}
is well established, it contains some novel elements
such as a
$\displaystyle{\left \{z_{cm}, A\right\}}$
boost and a
$\displaystyle{\left \{t_{cm}, A\right\}}$
rotation.
The timelike nature of the $A^{tz}$-dimension
stems from the opposite time/spacelike
assignments of $t$ and $z$, as expressed via
$\eta_{tt}\eta_{zz}=-1$, in accordance with
$\eta_{tt}\eta_{55}=-1$.
The accompanying energy/momentum relation
takes then the form
\begin{equation}
	m^2=E_{cm}^2+E_A^2-p_{cm}^2 ~,
\end{equation}
where the total energy has now two independent
sources $E^2=E_{cm}^2+E_A^2$.
The second energy operator 
\begin{equation}
	E_A=-2\pi i R\frac{\partial}{\partial A}
\end{equation}
is identified as the generator of loop area
expansions in the $\{t,z\}$-plane.
As a trivial consistency check one may verify
that a loop of a constant area, not necessarily
of a constant shape, would return the familiar
formulae describing a point particle moving in
$1+1$ dimensions.

%%%%%%%%%%%%%%%%%%%%
\subsection{From Special Relativity to Loop
Special Relativity}

The loop spacetime line element
eq.(\ref{newloopline}) and the subsequent
special cases discussed bring us one step
closer to our second goal, which is formulating
a Loop Special Relativity (LSR) theory capable
of supporting the Special Relativity (SR) limit
\begin{equation}
	\delta S^2=\eta_{ij}dx^i_{cm}dx^j_{cm}
	+O[\ell^2] ~.
	\label{SR2}
\end{equation}

The idea of trapping the  loop around
the compact fifth dimension can be conveniently
realized by executing the partial gauge choice
($\sigma$-redefinition)
\begin{equation}
	x^5(\tau,\sigma)=
	x_{cm}^5(\tau)+R\sigma ~.
	\label {x5tau}
\end{equation}
This way, starting from the most general Fourier
expansions for
\begin{equation}
	x^i (\tau,\sigma)=x^i_{cm} (\tau)+
	\ell \sum_{n\neq 0}\xi^i_{n} (\tau)e^{i n\sigma}~,
\end{equation}
we can significantly simplify the explicit expressions
for the associated infinitesimal projected areas.
Absorbing for simplicity the $\frac{\ell}{R}$ ratio 
within $\xi^i$, we find
\begin{eqnarray}
	&&\displaystyle{
	\frac{1}{2\pi R}\frac{dA^{i5}}{d\tau}}
	=\frac{d}{d\tau}x^i_{cm} ~, \\
	&&\displaystyle{
	\frac{1}{2\pi R}\frac{dA^{ij}}{d\tau}}
	=2R \frac{d}{d\tau}\sum_{n> 0}
	{\Im} (n~\xi ^i_n \xi^{j\star}_n) ~,
\end{eqnarray}
which can finally be substituted into
eq.(\ref{newloopline}).
Altogether, the LSR line element takes the
elegant form
\begin{equation}
	\boxed{
	ds^2_{LSR}=ds^2_{SR}-R^2(ds_E^2-ds_B^2)}
\end{equation}
Our anchor, the $3+1$ dimensional center of
mass SR line element, is hereby supplemented by
a $3+3$ dimensional Maxwell-like structure.
It should be noted that (i) All $O[\ell]$ mixed
pieces have been dropped away by the specific
gauge choice eq.(\ref{x5tau}), and consequently
that (ii) $x_{cm}^5(\tau)$ appears irrelevant as
it has no realization in our formalism.

Naturally, attention is devoted to the light cone
structure.
The special relativistic $ds^2_{SR}=0$ is replaced
by its loop special relativistic $ds^2_{LSR}=0$.
Thus, from the point of view of an observer,
unfamiliar with (or just insensitive to) LSR, as is
evident from
\begin{equation}
	ds^2_{LSR}=0\quad\Longrightarrow\quad
	ds^2_{SR}=R^2(ds_E^2-ds_B^2) ~,
\end{equation}
the center of mass light cone acquires
a 6-dim (hopefully Planck scale) Maxwell-like fine
structure.
Three of the extra dimensions are 'electric'
(timelike) and the other three are 'magnetic'
(spacelike).
They are supposed to fade away as $R\rightarrow 0$.
This seems to constitute the main physics fingerprint
of LSR.

%%%%%%%%%%%%%%%%%%%%
\subsection{Towards Loop General Relativity:
Nash global isometric Embedding}

Attempting to go beyond flatness, one would now
like to calculate the geodesic distance $L$ between
two loops which reside in a curved background, say
two parallel loops living on a 2-dim sphere of constant
radius $\ell$.
The crucial point is that the 2-sphere can be globally
embedded within a flat 3-dim space, that is
\begin{equation}
	\begin{array}{lll}
		& x(\theta,\sigma)=
		\ell\sin\theta \cos\sigma &  
		\vspace{3pt} \\
		&y(\theta,\sigma)=
		\ell\sin\theta \sin\sigma &
		 \vspace{3pt}  \\ 
		 &z(\theta,\sigma)=
		 \ell\cos\theta & 
	\end{array} ~,
\end{equation}
subject to the global constraint $x^2+y^2+z^2=\ell^2$.
The only non-vanishing projected area increment
is then
\begin{equation}
	\delta A^{xy}=
	2\pi \ell^2\sin\theta\cos\theta d\theta ~.
\end{equation}
Following our prescription, one invokes the Kaluza-Klein
topological term
\begin{equation}
	x_5(\theta,\sigma)=R\sigma ~,
\end{equation}
and consequently finds
\begin{equation}
	L=\frac{\ell}{R}\int_{\theta_1}^{\theta_2}
	\sqrt{\ell^2 \cos^2\theta+R^2}
	\sin\theta d\theta ~.
\end{equation}
However, from a variety of reasons outlined towards
the end of the introduction, this is not necessarily a
recipe for LGR. 
Still, one cannot rule out the possibility, make it a
conjecture, that Nash local/global embedding may
eventually play some role in constructing LGR.

\section{Is loop dynamics imperative?}

\subsection{Introducing Planck scale}

So far, we have demonstrated how to map a loop
event, residing in $M_n \otimes S_1$ Kaluza-Klein
spacetime, into a point-like event residing in a larger
$\frac{1}{2}n(n+1)$-dim flat loop spacetime whose
center of mass sub-metric is supplemented
by a Maxwell style higher dimensional companion.
We are also aware of the topological advantage that,
being wrapped around the fifth dimension, the loop
event cannot really shrink to a point-like event.
However,  as it stands, while the mandatory SR-limit
has been non-trivially recovered, the overall picture
is still not fully satisfactory.
The reasons are fourfold:

\smallskip
\noindent $\bullet$ Arbitrary loop shape: 
The loop-to-point mapping is unfortunately not
one-to-one, and only captures the projected
areas involved.
Eq.(\ref{newloopline}) is incapable of telling
one loop configuration from the other as
long as their projected areas are the same.
The challenge would be to convert such a
residual degree of freedom into a physically
tamed shape uncertainty.

\smallskip
\noindent $\bullet$ Unbounded loop size: 
Whereas the fifth dimension invoked is compact
by definition, characterized by its tiny Kaluza-Klein
radius $R$, the loop projected areas $A^{ij}$ can
in principle take arbitrarily large values.
In turn, unless the projected areas are themselves
of order $\mathcal{O}[R^2]$, the loop line element
$\delta S^2$ will be dominated by the Maxwell-like
term rather than by the center-of-mass term.

\smallskip
\noindent $\bullet$ Multiple timelike dimensions: 
Once $\eta_{ij}$ is specified, and $x^5$ is assigned
spacelike,
the signatures of Maxwell-like terms are not a
matter of choice.
To be specific, a single timelike spacetime
coordinate $t$ gives rise to $(n-1)$ timelike loop
spacetime  coordinates $A^{it}$ (and of course to
$\frac{1}{2}(n-1)(n-2)$ spacelike loop spacetime
coordinates $A^{ij}$).
This opens the door for problematic mathematical
as well as philosophical cause-and-effect issues,
arguing that the behavior of physical systems could
not be predicted reliably.
Saying this, note that several theories, F-theory
and 2T-theory \cite{2T} among them, do host multiple
timelike (and necessarily accompanied by spacelike)
dimensions.

\smallskip
\noindent $\bullet$ Multiple KK wrappings: 
Recalling that $\pi_1 (S_1)={\cal Z}$, the closed
loop event can carry an arbitrary integer winding
number $w=n\neq 0$.
We can of course always generalize our simple 
loop assumption, that is allow no self intersections
of the loop's oriented projections (which surround
the projected areas $A^{ij}$), but it is not natural
to do so in the presence of a non-trivial topology.
After all, a winding number $w=n\neq \pm 1$
loop gives rise to $(|n|-1)$ self intersections
on the KK cylinder.
A presumably quantum field theoretical self
interaction mechanism, capable of decomposing
a $w=n$ loop into $n$ separated $w=1$ simple
loops, is certainly in order, but unfortunately stays
beyond the scope of the present paper.

Appreciating the above potential drawbacks,
one is left with two apparently contradictable
options:

\noindent {\bf{Option I}}:
The exact configuration of the  loop event
does not show up at the classical level at all.
Had $R$ vanished as $\hbar\rightarrow 0$,
such an option would have been encouraged
by quantum mechanics, with every  loop
configuration carrying its own amplitude. 
In fact, this option favors the entrance of the
Planck scale into the game via
\begin{equation}
	R \sim \ell_P=
	\sqrt{\frac{\hbar G}{c^3}} ~,
	\label{Planck}
\end{equation}
implying that $R$ should
vanish at the $G\rightarrow 0$ limit as well
and of course as $c \rightarrow \infty$.

\noindent {\bf Option II:}
The exact configuration of the  loop event does
acquire a physical meaning already at the classical
level.
The introduction of dynamics via some string
theoretical action seems then unavoidable,
with the main goal being to encourage loop
events to shrink (positive string tension).
Such a dynamical approach would furthermore
account for the assumption that $A^{ij}\sim R^2$
and that the perimeter $2\pi R$ of the fifth
dimension share the one and the same length scale.

While the second option may seem easier to
utilize on technical grounds, it is the first
option which actually catches our imagination.
Thus, we choose to adopt the Planck length scale
eq.(\ref{Planck}), but without giving up the idea of
loop event self dynamics.
In other words, we attempt to make a compromise,
and choose

\noindent {\bf Option I+II}:
Translated into the Lagrangian formalism, we propose
\begin{equation}
	\boxed{
	{\cal I}={\cal I}_{LSR}+\Lambda {\cal I}_{NG}} 
	\label{ILambda}
\end{equation}
where ${\cal I}_{LSR}=\displaystyle{\int dS}$ is the
loop spacetime geodesic action, and
${\cal I}_{NG}=\displaystyle{\int \sqrt{-g_2}~d\tau d\sigma}$
stands for the familiar string theoretical Nambu-Goto
action (or alternatively for its Polyakov variant).
The first ingredient contains the SR-limit, but
is insensitive to the detailed structure of the loop.
The second ingredient does not have an SR-limit,
but forcefully governs the inner loop dynamics.
$\Lambda$ is a dimensionless coefficient, which
may be eventually elevated in some stage to the
level of of a Lagrange multiplier.
To show our point we discuss now in some detail the
simplest pedagogical case of sufficient complexity,
namely a 2-dim soap world-sheet embedded within
a 3-dim Euclidean space.

\subsection{No-go Nambu-Goto}

We now prove, as was claimed before, that the
Nambu-Goto action (as well as its Polyakov
vartiant) cannot consistently serve as a measure
of the 'distance' between two loops.
To stand on familiar geometrical grounds, we
choose to make our point using the simplest
pedagogical case of a 2-dim soap worldsheet
embedded within a 3-dim Euclidean space
\begin{equation}
	\left.
	\begin{array}{lll}
	x^1(z,\sigma)=r(z)\cos\sigma && \\
	x^2(z,\sigma)=r(z)\sin\sigma && \\
	x^3(z,\sigma)=z && 
	\end{array}
	\right. ~,
	\label{xiz}
\end{equation}
connecting two circles of equal radii
\begin{equation}
	r(-h)=r(h)=\ell~.
\end{equation}
The corresponding Nambu-Goto action reads
\begin{equation}
	I_{NG}=2\pi \int_{-h}^{h} 
	r(z) \sqrt{1+r^\prime(z)^2}~dz ~.
	\label{ING}
\end{equation}
The Euler-Lagrange equation and the
corresponding analytic solution are given by
\begin{equation}
	r r^{\prime\prime} -r^{\prime 2}-1=0 
	\quad\Longrightarrow \quad
	r(z)=\frac{\cosh k z}{\cosh k h} \ell~,
	\label{rNG}
\end{equation}
subject to the symmetric boundary condition
\begin{equation}
	\cosh kh=k\ell ~.
	\label{ka1}
\end{equation}
It can be numerically verified that there is no
solution for $\displaystyle{\frac{\ell}{h}< 1.508}$,
with the critical value marking a phase transition,
a phenomenon which can be experimentally
demonstrated with soap films.
The critical point is associated with the extra
mathematical condition
\begin{equation}
		\sinh kh=\frac{\ell}{h} ~.
\end{equation}
In other words, the Nambu-Goto action cannot be
interpreted as 'distance' between
loops separated too far (beyond criticality) apart.

\subsection{Kaluza-Klein modified Nambu-Goto}

We now show that the trick of adding a compact
5-th dimension and wrapping the loop event once
around it, that is
\begin{equation}
	\left.
	\begin{array}{lll}
	x^1(z,\sigma)=r(z)\cos\sigma && \\
	x^2(z,\sigma)=r(z)\sin\sigma && \\
	x^3(z,\sigma)=z && \\
	x^5(z,\sigma)=R\sigma &&
	\end{array}
	\right. ~,
	\label{xiz5}
\end{equation}
while having some advantage over the plain
NG case, is still quite problematic.
The former Nambu-Goto action eq.(\ref{ING})
gets now generalized into
\begin{equation}
	I_{NG+KK}=2\pi \int_{-h}^{h} 
	 \sqrt{(r(z)^2+R^2)(1+r^{\prime}(z)^2)}~dz ~,
	 \label{INGKK}
\end{equation}
and subsequently, the modified Euler-Lagrange
equation takes the form
\begin{equation}
	(r^2+R^2)r^{\prime\prime}
	-r(1+r^{\prime 2})=0 ~.
\end{equation}
The analytic solution looks very much like
the one given by eq.(\ref{rNG}), save for
the modified boundary condition
\begin{equation}
	\cosh kh=\frac{k\ell}{\sqrt{1-k^2 R^2}} ~,
	\label{ka2}
\end{equation}
replacing the former eq.(\ref{ka1}).

A critical $h_c= 2.790 R$ plays now a major
role.
For $h<h_c$, there is exactly one real solution
for every $a$.
For $h>h_c$, there can still be a single real
solution provided $\ell<\ell_{min}$ or $\ell>\ell_{max}$.
However, for $h>h_c$ and $\ell_{min}<\ell<\ell_{max}$, 
where $\ell_{min,max}$ are the local extrema
values of eq.(\ref{ka2}), there appear to be three
real solutions.
Not only is uniqueness lost, but furthermore,
one unexpectedly faces a hysteresis phenomenon.

There are, however, some encouraging news to
report on.
This has to do with the large loop separation
region $h\gg R$, a region which plain
${\cal I}_{NG}$ eq.(\ref{ING}) simply
could not reach.
In which case, we derive the long distance
behavior
\begin{equation}
	\frac{\cal I}{2\pi R} \simeq
	2h+ \frac{~\ell^2}{R}\tanh \frac{h}{R}+... ~,
\end{equation}
showing a small $\displaystyle{{\cal O}(\frac{~\ell^2}{R})}$
correction to the classical large value of $2h$.
%Revised%
For some classical wrapped string solutions see \cite{sol}.

\subsection{Combining LSR with Nambu-Goto}

\begin{figure}[h]
	\center
	\includegraphics[scale=0.27]{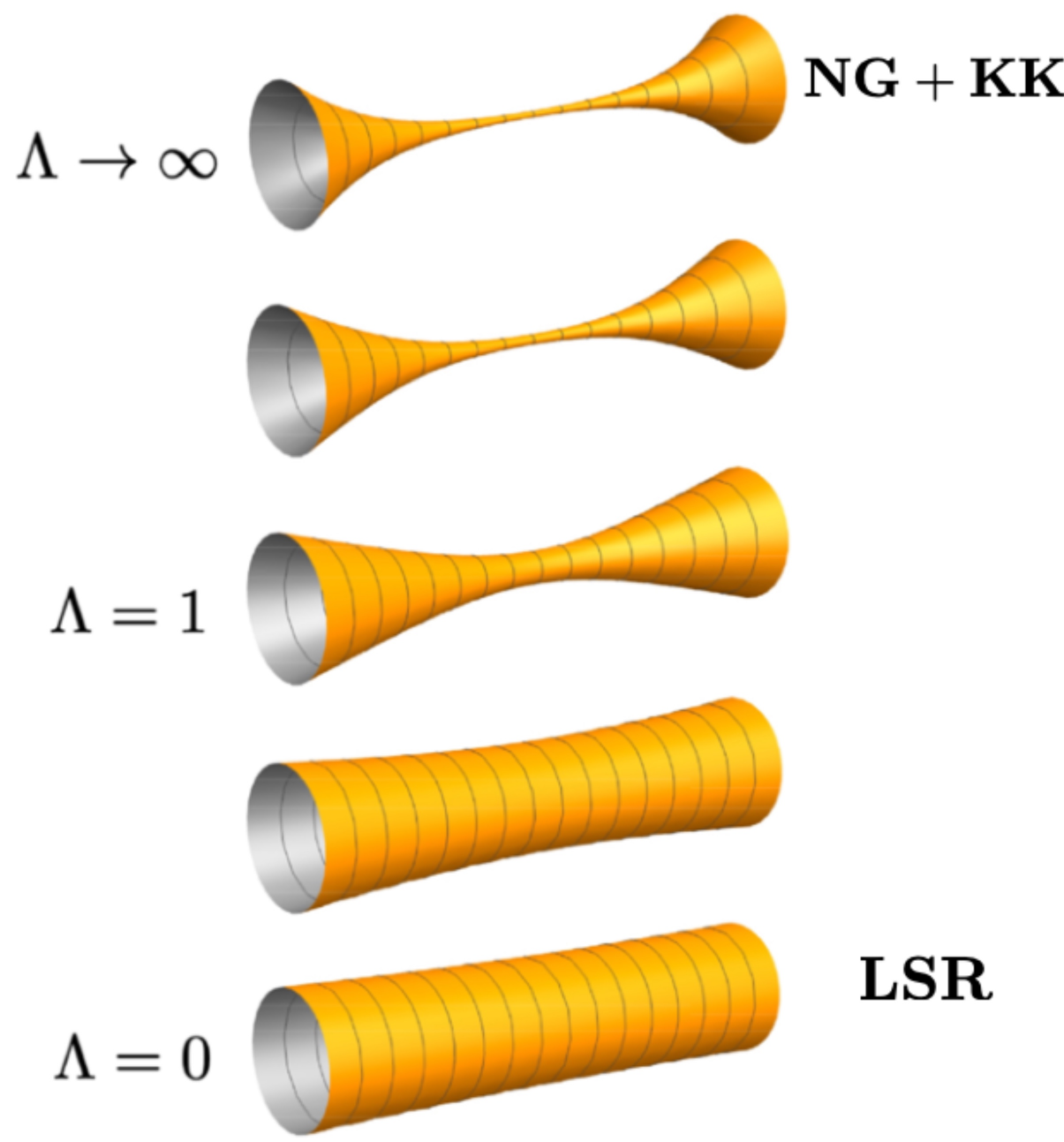}
	\caption{From LSR to NG+KK:
	Soap branes connecting two
	identical circular rings, residing in a 3-dim
	space and separated $2h$ apart.
	As $\Lambda$ grows, while holding the KK
	radius $R$ fixed, the LSR cylinder transforms
	into an NG+KK narrow waist candlestick
	(owing its stability to $R\neq 0$).}
	\label{fig4} 
\end{figure}

Sticking to eqs.(\ref{xiz5}), we now substitute
the various $x^i (z,\sigma)$ into the action
eq.(\ref{ILambda}) to arrive at
\begin{equation}
	{\cal I}_\Lambda=
	2\pi \int_{-h}^{h} \left(
	\sqrt{R^2+r^2 r^{\prime 2}}
	+\Lambda
	 \sqrt{(R^2+r^2)(1+r^{\prime 2})}
	 \right)dz ~.
	 \label{IL}
\end{equation}
Here, $\Lambda$ is just a dimensionless coefficient,
not a Lagrange multiplier.
$\Lambda \rightarrow \infty$ marks the NG-limit,
whereas $\Lambda \rightarrow 0$ takes us back
to the LSR territory (see Fig.\ref{fig4}).
At the first glance, the associated Euler-Lagrange
equation looks quite cumbersome, but once recasted
into the form $r^{\prime\prime}=f(r, r^\prime)$,
with $R$ and $h$ serving as parameters, it can be
numerically handled straight forwardly.

The numerical lesson is twofold:

\noindent (i) $\Lambda \geq 0$ on $z$-evolutionary
grounds, as otherwise we are necessarily driven into
an undesirable $r=0$ collapse.

\noindent (ii) $\Lambda \ll 1$ on self-consistency
grounds, as otherwise we loose track of LSR,
which becomes merely a perturbation on the
classical string action.
For such a small mixing parameter $\Lambda$,
we obtain
\begin{equation}
	\frac{{\cal I}_\Lambda}{2\pi R}
	=2h\left(1+\sqrt{1+\frac{\ell^2}{R^2}}
	\Lambda\right)+ ... 
	~\quad  \mathsf{for} ~\Lambda \ll 1 ~,
\end{equation}
to be fully contrasted with
\begin{equation}
	\frac{{\cal I}_\Lambda}{2\pi R}
	= 2h\Lambda \xi+...
	~\quad  \mathsf{for} ~\Lambda \gg 1
\end{equation}
where $\xi$ is some geometrical ${\cal O}[1]$ 
factor.

% Fig4  %%%

\section{Low frequency breathing mode} 

In a Lorentzian $M_4\otimes S_1$ background,
with the following loop assignments
\begin{equation}
	\left.
	\begin{array}{lll}
	x^0(\tau,\sigma)=\tau && \\
	x^1(\tau,\sigma)=r(\tau)\cos\sigma && \\
	x^2(\tau,\sigma)=r(\tau)\sin\sigma && \\
	x^3(\tau,\sigma)=0 && \\
	x^5(\tau,\sigma)=R\sigma &&
	\end{array}
	\right. 
	\label{txi5}
\end{equation}
corresponding to a circular loop evolving in the
$xy$-plane, the action Eq.(\ref{IL}) is traded for
\begin{equation}
	{\cal I}_\Lambda=
	2\pi \int \left(
	\sqrt{R^2-r^2 \dot{r}^2}
	+\Lambda
	 \sqrt{(R^2+r^2)(1-\dot{r}^2)}
	 \right)d\tau ~.
\end{equation}
Associated with the latter action is the Euler-Lagrange
equation
\begin{equation}
	\frac{\ddot{r}}{r}=
	-\frac{\dot{r}^2+\Lambda(1-\dot{r}^2)m(r,\dot{r})}
	{r^2+\Lambda(R^2+r^2)m(r,\dot{r})} ~,
	\label{EoM}
\end{equation}
where $m(r,\dot{r})$ is given explicitly by the ratio
\begin{equation}
	m(r,\dot{r})=
	\frac{(R^2 -r^2 \dot{r}^2)^\frac{3}{2}}
	{R^2 (R^2+r^2)^\frac{1}{2}
	(1-\dot{r}^2)^\frac{3}{2}} ~.
	\label{m}
\end{equation}
Without loosing generality, the accompanying
initial conditions are $r(0)=\ell$, $\dot{r}(0)=0$.

The solutions $r(\tau)$ of eq.(\ref{EoM}) are
in general oscillatory in $\tau$, with frequencies
of the general form
\begin{equation}
	\omega_\Lambda=
	\frac{1}{R}f_\Lambda (\frac{\ell}{R})
\end{equation}
which we now proceed to extract.
No oscillatory solutions, however, for
\begin{equation}
	-\frac{\ell^2}{R\sqrt{R^2+\ell^2}}
	\leq \Lambda \leq 0 ~,
\end{equation}
an important observation when the small
$\Lambda$ regime is one's preference.

A remark  is in order.
The fact that the oscillatory solutions pass
through $r=0$, corresponding to an orientation
flip of the loop in the $xy$-plane, is of no
special concern.
Owing to the underlying topology, that is
being wrapped around the Kaluza-Klein
cylinder, the loop cannot really shrink to
a singular point.

Three trivial cases can be immediately verified:

\noindent $\bullet$ $ \Lambda = 0$ marks the
LSR limit.
As explained earlier, the area $\sim r^2$ evolves
linearly with $\tau$, and given our initial
conditions, we face
\begin{equation}
	r(\tau)=a ~.
\end{equation}
In light of the forthcoming analysis, we assign
\begin{equation}
	\omega_{LSR}=0 ~.
\end{equation}

\noindent  $\bullet$ $\Lambda \rightarrow \infty$  
is recognized as the NG+KK limit.
In which case, we recover radial oscillations via
$r(\tau)=\ell \cos \omega\tau$, with frequency
\begin{equation}
	\omega_{NG}=\frac{1}{\sqrt{R^2+\ell^2}} ~.
	\label{1/Ra}
\end{equation}

\noindent $\bullet$ The limit
$ \displaystyle \frac{\ell}{R}\rightarrow 0$ is
trustfully translated into
$ \displaystyle{\frac{r(\tau)}{R}\rightarrow 0}$.
As the scale of $r(\tau)$ shrinks away,
one immediately notices that $m(r,\dot{r})\rightarrow 1$.
Altogether, as could have been expected, and for
any finite $\Lambda$, we are back to NG+KK only with
\begin{equation}
	\omega_{\Lambda} \rightarrow \frac{1}{R}~.
	\label{1/R}
\end{equation}

\begin{figure}[h]
	\center
	\includegraphics[scale=0.33]{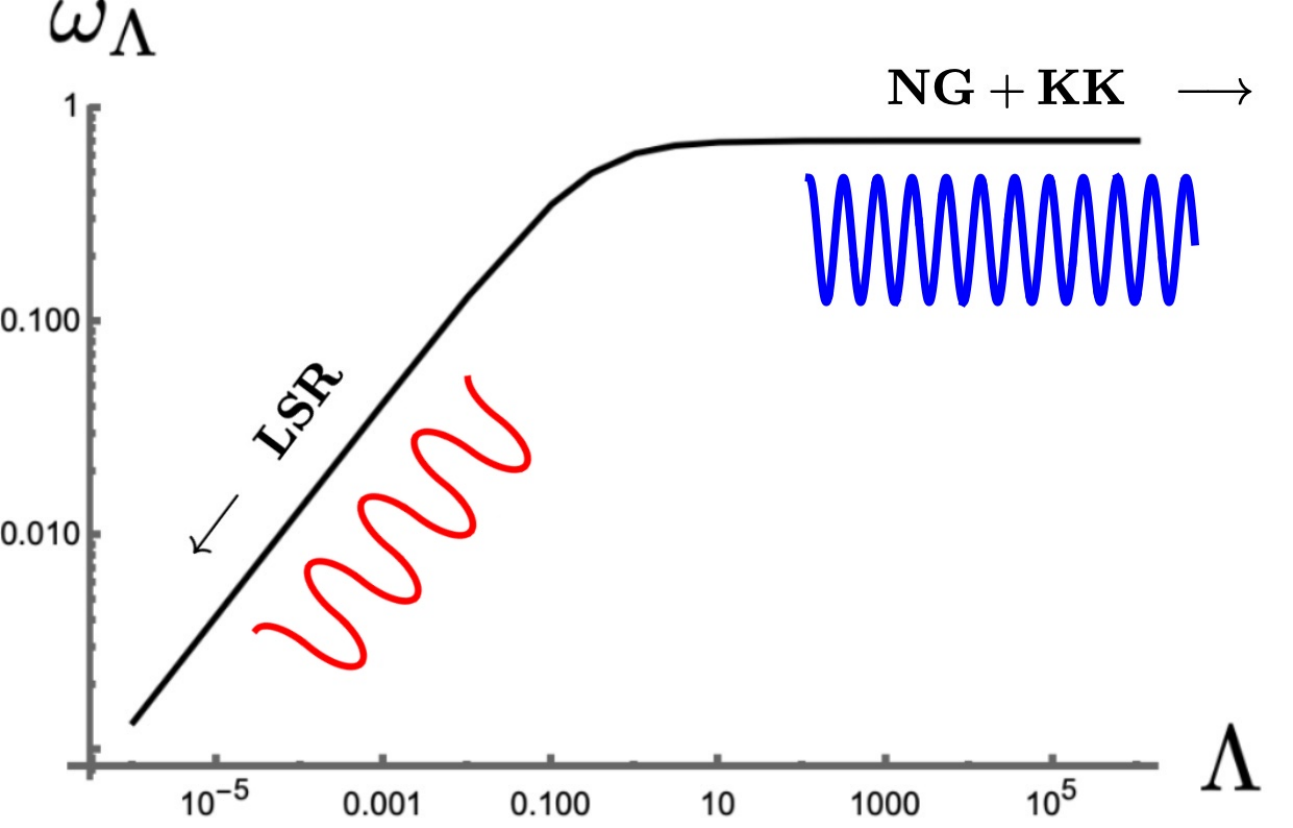}
	\caption{High to law frequency transition
	as a function of the string tension $\Lambda$
	(plotted for the special case $\ell=R=1$).
	While $\omega_\Lambda$ is
	$\Lambda$-independent at the high-$\omega$
	regime, it is $\sim \sqrt{\Lambda}$
	at the low-$\omega$ regime}
	\label{h2l} 
\end{figure}

For a positive string tension $\Lambda >0$,
we now prove the existence of a periodic breathing
mode, and attempt to study the entire range
$\displaystyle{\omega_{LSR}< \omega_\Lambda
< \omega_{NG}}$.
Our interest lies however with the low frequency mode
associated with the LSR governed case $\Lambda \ll 1$.
The transition from the low to high frequency regimes
is depicted (for the special case $\ell=R$) on a log-log
graph, see fig.(\ref{h2l}).
% Fig5  %%%

\noindent $\bullet$ $\Lambda \ll 1$ is our case
of interest. 
On simplicity and pedagogical grounds, we choose
to make our points using the prototype case $\ell=R$.
The periodic numerical solution then suggests
$\displaystyle{\omega_\Lambda\sim
\frac{\sqrt{\Lambda}}{R}}$,
a result which we now extract semi-analytically.
We start by noticing that $m(\tau)$ is a shallow
function of $\tau$.
It starts at $m(0)=1/\sqrt{2}$ and stays below $1$
up to an extremely narrow yet finite peak when
$r(\tau)$ passes through zero.
We then replace $m(\tau)$ by its $\tau$-average
constant value $\frac{1}{\sqrt{2}}< m < 1$ (a tenable
value is $m\simeq0.8$). 
And finally, throwing away two terms of tiny
numerical contributions, we arrive at the
approximated equation of motion
\begin{equation}
	\frac{\ddot{r}}{r}\simeq
	-\frac{\dot{r}^2+\Lambda m}
	{r^2+R^2 \Lambda m} ~.
\end{equation}
Not only does the latter equation admit an
analytic solution, but it forcefully captures the
essence (shape, extrema, zeroes) of the original
numerical solution of eq.(\ref{EoM}).
The semi-analytic solution is given explicitly by
\begin{equation}
	\frac{\tau}{R}=E\left [\frac{\pi}{2},
	\frac{-1}{\Lambda m}\right]
	-E\left[\arcsin\frac{r(\tau)}{R},
	\frac{-1}{\Lambda m}\right]~,
	\label{sol}
\end{equation}
where
$E[\phi,\chi]\equiv \int_0^\phi \sqrt{1-\chi \sin^2
\theta}~d\theta $ stands for the elliptic integral of
the second kind.

The role of the constant term in eq.(\ref{sol}) is to
reassure that $r(0)=R$.
Furthermore, the radial velocity
\begin{equation}
	\dot{r}(\tau)\simeq -\frac
	{\displaystyle{\sqrt{1-\frac{r^2}{R^2}}}}
	{\displaystyle{\sqrt{1+\frac{r^2}{R^2 \Lambda m}}}}
\end{equation}
not only confirms that $\dot{r}(0)=0$, but also
correctly takes care of the physical upper bound
$|\dot{r}(\tau)| \leq 1$.
Our main bonus now being the frequency formula
\begin{equation}
	\omega_\Lambda=\frac{2\pi}{\Delta \tau}=
	\frac{\pi}{2R\displaystyle{E\left[\frac{\pi}{2},
	\frac{-1}{\Lambda m}\right]}}~.
\end{equation}
Appreciating the fact that $\displaystyle{E[\frac{\pi}{2},\frac{-1}{x}]}$
behaves like $\displaystyle{\frac{1}{\sqrt{x}}}$ for $x\ll 1$,
we end up with the low frequency limit
\begin{equation}
	\boxed{
	\omega_\Lambda\simeq
	\frac{\pi\sqrt{\Lambda m}}{2R}
	\quad~ \mathsf{for}~\Lambda \ll 1 }
\end{equation}
Note that for $\Lambda\gg 1$ we recapture
eq.(\ref{1/R}).

\section{Epilogue} 

This paper focuses on an unexpected role played
by the compact fifth dimension, serving as the missing
topological ingredient which allows for the conversion
of an area metric into a legitimate line element in loop
spacetime.
Ironically, it is the Kaluza-Klein ansatz
which paves the way for Loop Special Relativity
(LSR) to exhibit the indispensable Special Relativity
(SR) limit.
Associated with $M_4 \otimes S_1$ is then a
10-dim loop spacetime metric, whose 4-dim
center-of-mass core term is supplemented by a
6-dim Maxwell-style Planck-scale fine structure
which drops away at the shrinking loop limit.

There are, however, as listed in Chapter 4, four
flies in the LSR ointment, so to speak.
Being kinematical in nature, LSR suffers from
arbitrary loop shapes, unbounded loop sizes, multiple
timelike dimensions, and multiple KK wrappings.
Some of these problems require a dynamical solution,
suggesting the presence of a positive string tension.
But dynamical Nambu-Goto (or Polyakov) by itself
will not do either, as it clearly lacks the SR limit in
the technical sense of eq.(\ref{SRlimit}).
The action marriage eq.(\ref{ILambda}) is perhaps
an elegant, and by far the simplest, way out.
It comes with a clear signature, that is a
low-frequency breathing mode.
And finally, just a reminder that the generalization
of LSR (Loop Special Relativity) into LGR (Loop
General Relativity) is still in order.
Nash global isometric embedding may play a key
role in constructing the latter theory.

\begin{acknowledgments}
The idea presented here was shaped
up back at 1996, during a visit to N.K. Nielsen
(University of Southern Denmark), and has later
served to seed the M.Sc. thesis of Nadav Barkai
(Ben Gurion University, 2007, unpublished) which
dealt with "Gauge invariance of the second type".
\end{acknowledgments}

% \bibliography{Citations}

% Here we use the simpler non-BibTeX alternative

\end{document}